\documentclass[%
 preprint,
 amsmath,amssymb,
]{revtex4-1}

\usepackage{graphicx}
\usepackage{amssymb}
\usepackage{setspace}
\usepackage[utf8]{inputenc}
\usepackage{array}
\usepackage{natbib}
\usepackage{amsmath}
\usepackage{multirow}
\usepackage{hyperref}
\usepackage{subfigure}

\DeclareMathOperator*{\argmax}{arg\,max}

\setcitestyle{authoryear,open={(},close={)}}

\usepackage{color}
\definecolor{redcolor}{rgb}{1.0,0.,0.}

\definecolor{bluecolor}{rgb}{0,0.,1}

\begin{document}

\preprint{}

\title{{Analyzing the influence of prolific collaborations on authors productivity and visibility}}

\author{Ana C. M. Brito$^1$,  Filipi N. Silva$^2$ and Diego R. Amancio$^1$}

\affiliation{$^1$Institute of Mathematics and Computer Science, University of S\~ao Paulo, S\~ao Carlos, Brazil\\
$^2$Indiana University Network Science Institute, Bloomington, Indiana 47408, USA\\
}

\date{\today}

\begin{abstract}
%
%
%
Science has become more collaborative over the past years, a phenomenon that is related to the increase in the number of authors per paper and the emergence of interdisciplinary works featuring specialists of different fields. In such a environment, it is not trivial to quantify the individual impact of researchers. Here we analyze how the most prolific collaboration tie (in terms of co-produced works) of an established researcher influences their productivity and visibility metrics. In particular, we check how the number of produced works, citations and h-index rank of an researcher changes when their works also coauthored by their prolific collaborators are not considered. We observed different patterns of prolific collaborator influence across the major fields of knowledge. More specifically, in formal and applied sciences, the prolific collaborators seem to play an important role to the visibility metrics of authors even when they are among the highly cited. Such a results can help stakeholders to better understand the collaboration patterns and draw measures of success that also consider collaboration ties. 
\end{abstract}

\maketitle

\section{Introduction}

Researchers are looking for collaborations more than ever~\citep{beaver2001reflections, freeman20151,lariviere2015team,viana2013time}. Over the past few decades, team sizes have been increasing across many research fields in science, with scholars seeking collaborations to increase productivity, enrich knowledge by sharing experiences with colleagues, access new resources, attain higher impact, among many other reasons~\citep{bukvova2010studying, abramo2009research}. Projects bridging two or more fields are usually undertaken by larger teams with researchers from different disciplines. However, this is a relatively recent notion in Science and may not have spread uniformly across different disciplines as some areas incorporate collaborations in the research process most naturally than others~\citep{qin1997types}.

Collaborations can emerge in different contexts. Sometimes, it happens between graduate students, researchers affiliated to the same institutions, supervisors and supervised, etc.  In general, it is expected that authors have few collaborators at the beginning of their academic life, working predominantly with their supervisors more than other researchers. At the same time, different research fields may have different forms of measuring success (e.g., through productivity, citations, etc)~\citep{abramo2013individual,abramo2014you}. In addition to that, even though the use of similar metrics, such a task can be challenging once different disciplines also present different dynamics on how authors collaborate and beginning of the author's academic career.

In the literature, we can find various approaches aiming to quantify scholars' academic productivity, contribution and impact~\citep{garcia2012extension, abramo2014you, fenner2014altmetrics,correa2017patterns,brito2021associations}. However, these characterizations rarely take into account the authors' collaborations nuances, for example, how distributed are the citations across different team sizes. It is not trivial to quantify the individual impact on authors' success given such a variety of patterns and scales of collaborations. In this work, we aim to characterize authors productivity taking into account their collaborations, and compare these patterns across different disciplines. 

We focus our analyses on understanding the effect of collaboration ties on different fields of study for established authors. In particular for researchers that already published at least a minimum number of papers. We consider the co-authorship of papers as a proxy for collaboration ties between authors. The analysis is then performed by means of author-level success metrics grouped by research fields of study. We use metrics such as the h-index, the number of citations, the number of publications and verify how they vary with the presence or absence of their prolific collaborator (or top collaborator). In this context, the top collaborator is the collaborator of an author whose coauthored papers received more citations. Our analysis is guided by the following research questions:
\begin{enumerate}
    \item How does the collaborations strength change in different fields of study?
    \item What is the impact of the top collaborator on authors' productivity?
\end{enumerate}
%


We found that the top collaborator influences to authors metrics have different patterns depending on their fields of study. In particular, three main patterns were observed across disciplines. First, humanities disciplines were found to have authors predominantly publishing single-authored papers or with a few collaborators. In this case, the top collaborator does not seem to have a strong influence on individual metrics. The second type of pattern was observed mostly for natural and applied disciplines, with authors having many collaborators but  with the top collaborator also not influencing much on their success metrics. The third group of patterns is composed of formal and applied sciences. These  disciplines display authors metrics highly influenced by their top collaborators. Interestingly, these tendencies are different for the specific case of highly cited authors since their success metrics are not highly influenced by their top collaborators. This pattern was evident across all studied fields of study.
%
Our results may shed light on the understanding of the importance of collaborations to researchers' visibility. 


\section{Related works}
\label{sec:related}


\subsection{Comparing productivity in different disciplines}

In order to measure individual contributions, \cite{batista2006possible} normalized the traditional h-index by the average number of authors co-authoring papers in the h-set. The study compared Brazilian researchers in four disciplines, namely Physics, Chemistry, Biology and Mathematics. 
Lists comprising the top 10 Brazilian authors were generated according to the traditional and the proposed normalized h-index. The overlap between these two lists was found to be field dependent. The highest overlap occurred in Mathematics (90\%), which indicates that the normalization barely impacts the rank. Conversely, Physics presented the lowest overlap value, 10\%. The results suggest that the consideration of the number of collaborators can highly impact the ranking of authors. \cite{batista2006possible} advocates, however, that the proposed normalization allows comparing h-indexes across fields.

\cite{ajiferuke2010citer} and \cite{ajiferuke2010comparison} proposed the use of \emph{citers} instead of citations to quantify academic influence. The idea is to count the number of individuals influenced by the author's work. 
In addition to using traditional citation metrics such as citation counts, they also proposed the ch-index. This measurement is analogous to the traditional h-index.  But instead of counting citations, the authors count the number of \emph{citers} in publications. Therefore, ch-index = $x$ if $x$ is the highest integer for an author with $x$ publications that are cited by at least $x$ different \emph{citers}. 
Significant correlations were found between the ch-index and  traditional citations metrics. However,  the ch-index seems to play a complementary and relevant role to investigate \emph{citers} and citations variations in different fields or to compare individuals with lower citation counts. 

\cite{ajiferuke2010comparison} extended the study of \cite{ajiferuke2010citer} in order to compare \emph{citer} patterns across disciplines. In addition to the already mentioned metrics, they introduced the \emph{reciter} rate, which is defined as the number of unique authors citing a paper in a recurrent way. \cite{ajiferuke2010comparison}  analyzed 90 highly cited authors from three different disciplines: Social Sciences, Mathematical/Engineering Sciences, and Biological/Medical Sciences. In general, \emph{citer} and citation measurements were found to be correlated. However, some different patterns were found in different disciplines. 
In Social Sciences, the number of authors per paper and reciter rates are lower than the other fields. The authors conclude that \emph{citer}-based metrics seem to be useful when the number of collaborations in the field is low. Conversely, in Mathematical, Engineering, Biological and Medical Sciences, {\emph{citer} and \emph{citation} metrics may present distinct values, even though they are correlated. 
}
%
%
Visibility analysis based on \emph{citer/reciter} counts also possibilities to identify if a large number of citations corresponds to large number of authors being influenced. 

\cite{ioannidis2016multiple} proposed a composition of indexes to quantify authors visibility. Their analysis a large list of citations metrics: number of citations, number of citations for the papers single-authored or first author, citations for single, first or last author, h-index and Schreiber co-authorship adjusted Hm-index. The combination of these indexes, referred to as Composite Score, was to used to 
they identify the top $1000$ researchers in the considered dataset. 
The authors found that an expressive number of notorious authors appear in their list while being absent in the list of top cited researchers. \cite{ioannidis2020updated} provided an updated version of the list of most influential researchers according to the Composite Score. 

\subsection{Collaborations and productivity}

\cite{abramo2009research} advocates that the relationship between collaboration and productivity is not trivial. The authors studied the 
Italian academic system across eight disciplines. To measure the productivity they used the total number of papers. 
{The relevance (referred to as quality index) of the research conducted by a group of researchers was computed by considering not only the number of papers published, but also journals impact factor.} 
The study found that the way collaborations influence productivity depends  on the research discipline. A strong correlation was observed for industrial and information engineering. In both mathematics and computer science fields, a correlation {between collaboration intensity and the average quality of the publications was found.} 
{For most of the disciplines investigated, the average quality {index} turned out to be positively affected by inter-university collaborations.}
This work also found that interdisciplinary fields have usually display a more collaborative behavior.

\cite{lariviere2015team} investigated papers published in two broad fields of study: natural/medical sciences and social sciences/humanities. They studied collaboration at the author level. Inter-institutional and country collaborations were also investigated. The results showed that all three types of collaborations are increasing in both fields. Single authored papers were found to be less usual, mainly in the natural and medical sciences. Most of the papers were found to be result of inter-institutional collaborations. 
Most importantly, \cite{lariviere2015team} showed that scientific impact seems to be influenced by the number of authors of the papers. They found no correlation between the diversity of authors country and impact.

\cite{abramo2015relationship} conducted a similar investigation for the Italian case, most of their findings confirmed results the results obtained by \cite{lariviere2015team}. The main conclusions pointed to the positive correlation between the number of authors of the papers and the number of citations received. Surprisingly, the correlating was even stronger for Social Sciences, Art, and Humanities.

\section{Methodology}
\label{sec:meth}

\subsection{Dataset and fields of study}

The Microsoft Academic Graph (MAG) is a large dataset containing scientific publications, citations, publications metadata such as authors and their institutions, journals, conferences, and fields of study~\citep{wang2020microsoft}. In this paper, we use papers published between 1950 and 2020. Because we are mostly interested in analyzing the collaborative patterns of mid-career and senior researchers, we considered only authors with at least $10$ papers and 200 citations in the dataset. We also disregarded papers co-authored by more than 10 authors, since the actual collaboration between authors in these cases is not trivial to measure. This same procedure has been applied in similar studies on collaboration networks~\citep{li2019reciprocity}.  
{Figure \ref{fig:num-authors} of the Supplementary Information (SI) shows the distribution of the number of authors per publication.} After this filtering step, the resulting number of authors and papers was 2,630,275 and 243,030,343; respectively. 

%
%



\begin{figure}[ht!]
    \centering
    \includegraphics[width=0.7\textwidth]{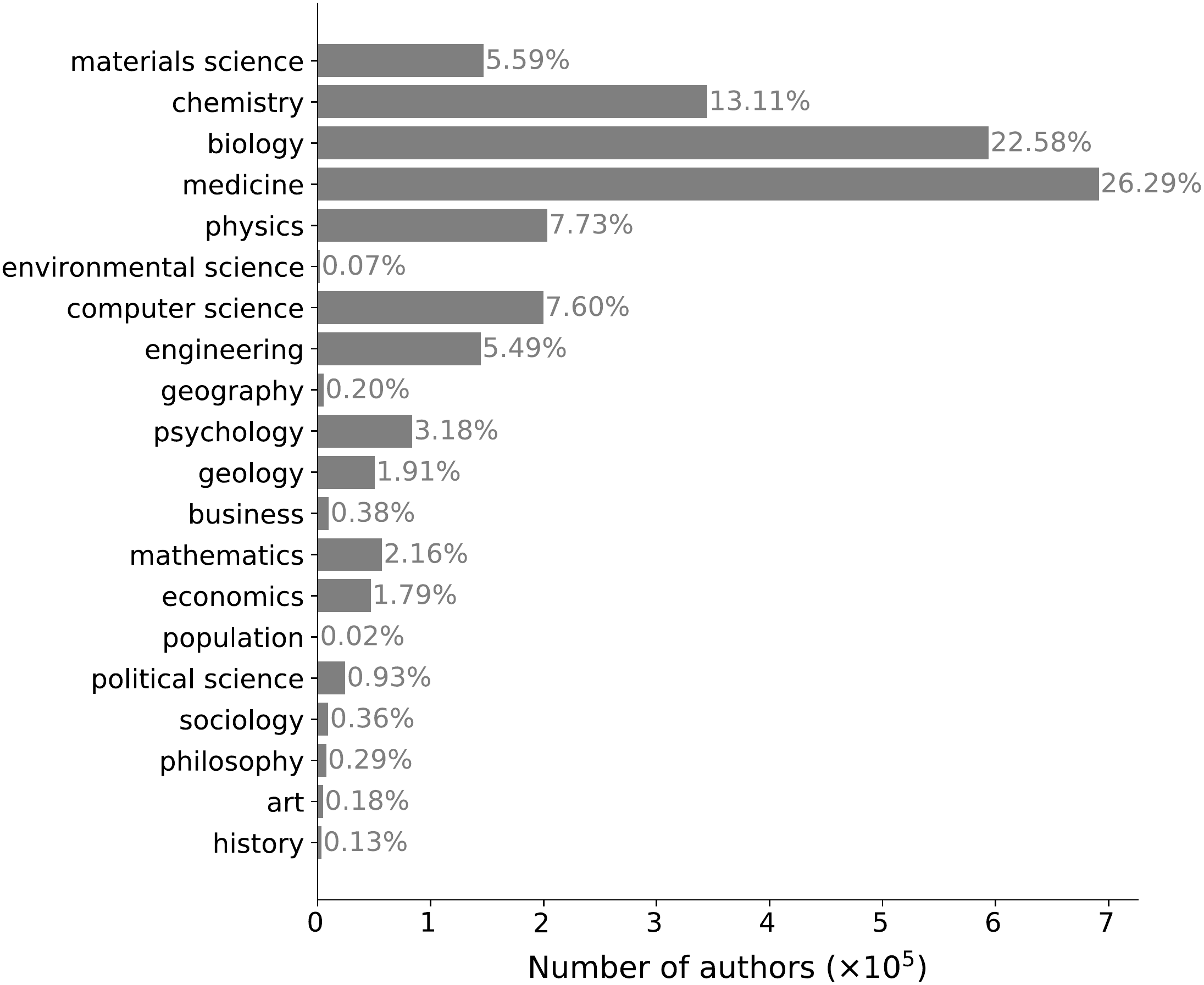}
    \caption{Number of authors  per field of study considering papers published between 1950 and 2020. We also show after each bar the corresponding fraction of authors in the respective field. 
    }
    \label{fig:authorsdist}
\end{figure}

Because we are also interested in collaboration patterns arising from diverse fields, one must first classify authors in fields of study. Here we employ the MAG fields of study, which are defined for each paper. Fields and subfields are organized in a hierarchy represented by a directed acyclic graph. However, because some subfields can be too specific, we did not use them directly. For example, the topic \emph{Katz centrality} is a child of \emph{Network science}, which in turn is a child of \emph{Complex Networks}. Instead, we map papers to fields at a hierarchy level corresponding to the major knowledge fields, such as Mathematics, Computer Science and Physics. This is accomplished by first mapping all the subfields to \emph{top fields} by walking upwards (towards the root) across the hierarchy until we reach a field of study at the desired level (major areas of knowledge). For instance, papers associated to the \emph{Katz centrality} field of study, are mapped both to \emph{Mathematics} and \emph{Computer Science}. A list of fields considered in this work is illustrated in Figure \ref{fig:authorsdist}. To account for multiple fields of study associated to a paper (and subfields associated to multiple top fields) we also employ a weight $w_P(f_{top})$ indicating the relevance of the a top field of study $f_{top}$ to a paper $P$, which is computed as:

\begin{equation}
    w_P(f_{top}) = \frac{1}{|F_P|} \sum_{f\,\in\,F_P} \frac{|\textrm{Parents}(f) \cap \{f_{top}\}|}{|\textrm{Parents}(f)|},
\end{equation} 
%
where $F_P$ is the set of subfields of paper $P$ and $\textrm{Parents}(f)$ is the set of top fields obtained after mapping a subfield $f$ upwards across the hierarchy.

The next step is assigning fields for authors based on their published papers. Here we considered each paper contribution as having the same relevance to define the authors field (and respective weight). Thus the weight $w_A'(f_{top})$ associating an author $A$ to a top field $f_{top}$ can be given by:
\begin{equation}
w_A'(f_{top}) = \sum_{p\,\in\,P_A} w_p(f_{top}),
\end{equation}
where $P_A$ is the list of papers published by author $A$.

Finally, the field of study of an author $A$ is the field of study with the highest weight: 
\begin{equation}
F(A) = \argmax_{f_{top}}~w_A'(f_{top}).
\end{equation}



The obtained distribution of fields of study at the author-level is shown in Figure \ref{fig:authorsdist}. As expected, fields size are diverse. While Medicine and Biology comprises almost half of all authors, Environmental Sciences encompasses less than $0.1\%$ of the universe of considered researchers. 

\subsection{Network of influencies}


We propose a directed and weighted network to quantify collaborations importance. All nodes and edges are established as in a traditional collaboration network~\citep{amancio2015topological,amancio2012use,pelacho2021analysis,viana2013time}. Two authors are linked if they co-authored at least one paper.  We now aim at quantifying for an author A how important is the collaborator B. Here we \emph{illustrate} the importance of collaborations in terms of citations of scientific works accrued in the collaboration. As we shall show, any relevance index can be used to define this weight.
If $c_{AB}$ is the number of citations accrued by papers co-authored by $A$ and $B$, and $c_A$ is the number of citations received by $A$, then the weight $w_{AB}$ is computed as 
\begin{equation} \label{eq:besos}
    w_{AB} = \frac{c_{AB}}{c_A}.
\end{equation}
Note that $w_{AB}$ can be interpreted as the importance of the collaboration between $A$ and $B$ for researcher $A$. Note that $w_{AB}$ may not be equal to $w_{BA}$. 
For this reason, we are given a directed network of influence. In our analysis, we assign the weight $w_{AB}$ for the edge with $A$ and $B$ and source and target nodes, respectively. Thus, 
out-going edge weights are normalized, i.e. 
%
$\sum_X w_{AX} = 1,$
%
where $X$ is co-author of $A$. The \emph{top-collaborator} of $A$, $T(A)$, is defined as:
\begin{equation} \label{eq:topa}
    T(A) = \argmax_X w_{AX}.
\end{equation}
%

\begin{figure}[h]
    \centering
    \includegraphics[width=0.7\textwidth]{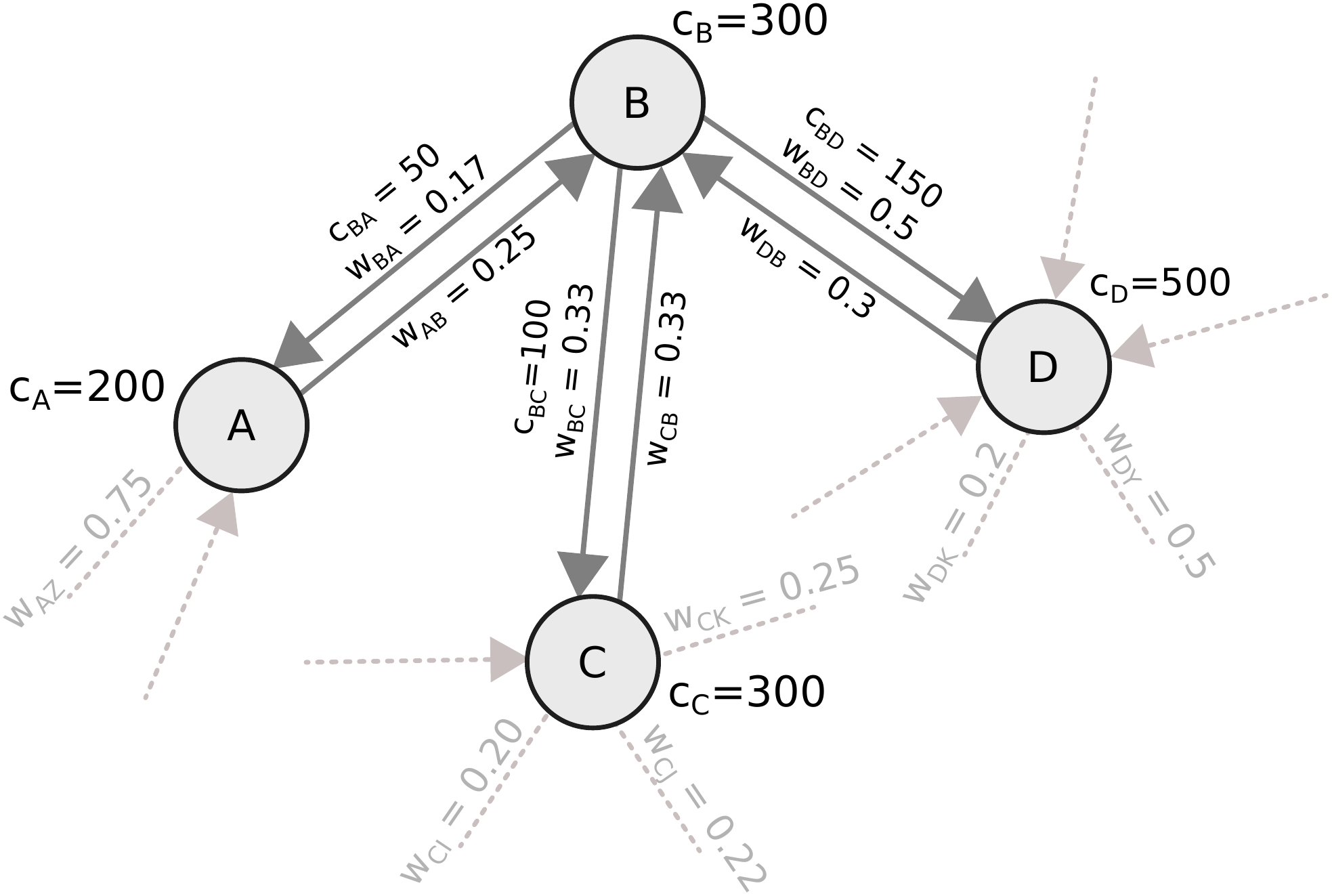}
    \caption{A toy example of a network of influences. Author A and B received $c_A = 200$ and $c_B = 300$ citations. Considering only the set of paper co-authored by $A$ and $B$, they received $c_{AB} = 50$ citations. For this reason, the importance of $B$ for author $A$ is $w_{AB} = 50/200 = 0.25$. Analogously, the importance of $B$ for $A$ is $w_{BA} = 50/300 = 0.17$.  }
    \label{fig:toynet}
\end{figure}

Figure \ref{fig:toynet} illustrates an example of the network of influence with edges weights defined according to equation \ref{eq:besos}. $A$ and $B$ accrued respectively $c_A = 200$ and $c_B = 300$ citations. Papers co-authored by $A$ and $B$ received $c_{AB} = 50$ citations. Taking $A$ as reference, the importance of $B$ is $w_{AB} = 0.25$, i.e. $B$ contributes to $25\%$ of all citations received by $A$. Note that, in this example, $A$ contributes to only 17\% of the citations received by $B$, i.e. $w_{BA} = 0.17$. $B$ is more benefited from collaborations with $D$, since 50\% of citations to B comes from papers co-authored with $D$, i.e. $w_{BD} = 0.50$. 
%

\begin{figure}[h]
    \centering
    \includegraphics[width=0.55\textwidth]{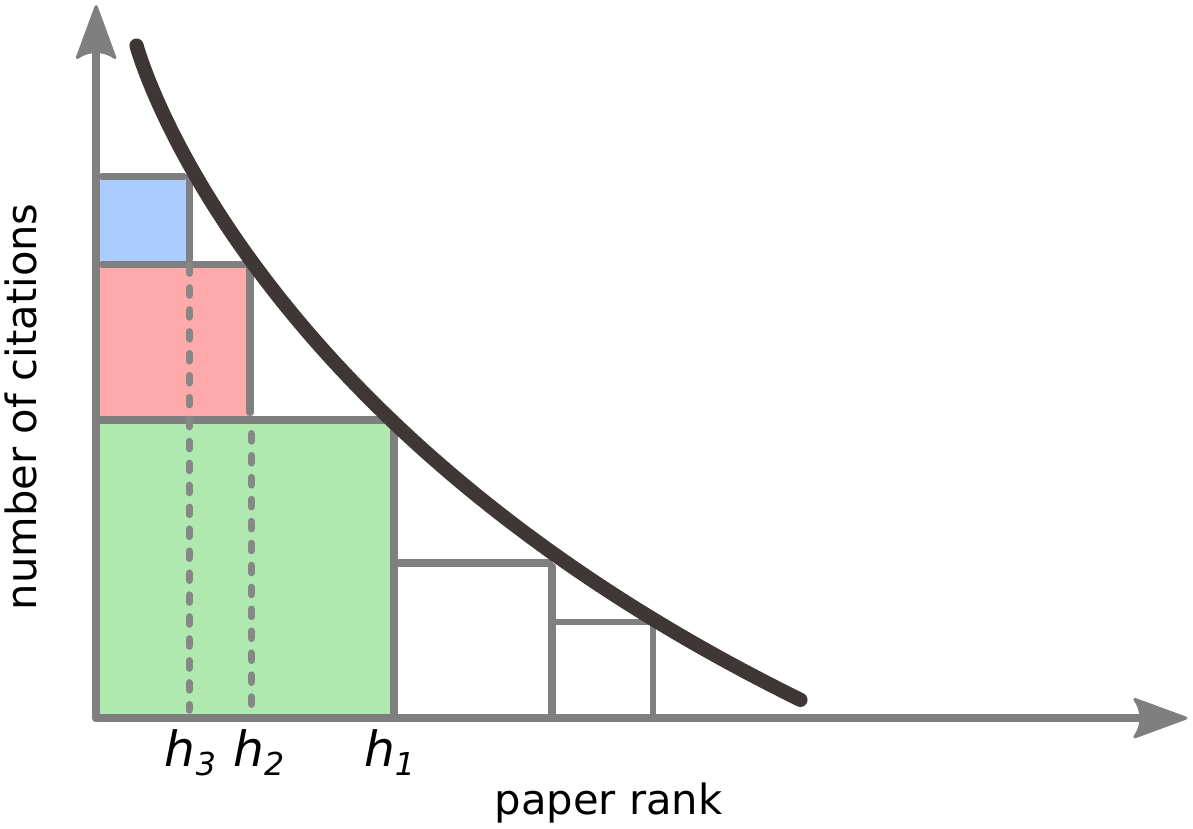}
    \caption{Extended h-index example. The traditional h-index ($h_1$) is computed by using all set of papers (green square). $h_1$ is the side length of the largest square fitting within the distribution. The square has its left down vertex located at the position $(0,0)$. 
    A recursive idea is used to computed $h_2$ and $h_3$. To compute $h_2$, we detect the side length of the largest square fitting within the distribution (see red square). In this case, the the considered square has its left down vertex located at the position $(0,h_1)$. The same procedure is applied to compute $h_3$, but now with the left down vertex located at the position $(0, h_1+h_2)$. }
    \label{fig:extendedhindex}
\end{figure}

In addition to the number of citations, we also computed the weight in equation \ref{eq:besos} in terms of the productivity (i.e. number of published papers). Both productivity and visibility were also considered by using the h-index. Because we are interested in ranking authors, the traditional h-index is not enough because many authors may share the same h-index value. For this reason, 
we used an extended version of the h-index that considers the distribution of citations in the $h$-set~\citep{garcia2012extension}. The extended version of the h-index ($h^{(E)}$) is computed as a sequence of $h$'s, i.e.  $h^{(E)} = \{h_1, h_2,h_3\ldots\}$. To compute $h^{(E)}$, papers are sorted in decreasing order of citations, as shown in Figure \ref{fig:extendedhindex}. $h_1$ corresponds to the traditional h-index. $h_2$ denotes the $h$-index computed in the $h_1$ set, but considering that every paper in the $h_1$ set has $h_1$ less citations than it actually has. This process in then recursively applied to compute $h_2, h_3\ldots$. A graphical illustration of this recursive process is illustrated in Figure \ref{fig:extendedhindex}.   

\section{Results} \label{sec:results}

\subsection{Top-collaborator influence across fields}

Here we aim at understanding the importance of top collaborators in authors productivity and visibility. Because some patterns may arise due to the different characteristics of the disciplines, we first analyze the main features of our dataset to better understand if the patterns we find can be trivially explained by simple subfield characteristics. For this reason, we analyze the distribution of the number of citations, number of papers and authors' birth year of each subfield.

\begin{figure}[h]
  \centering
  \includegraphics[width=0.9\textwidth]{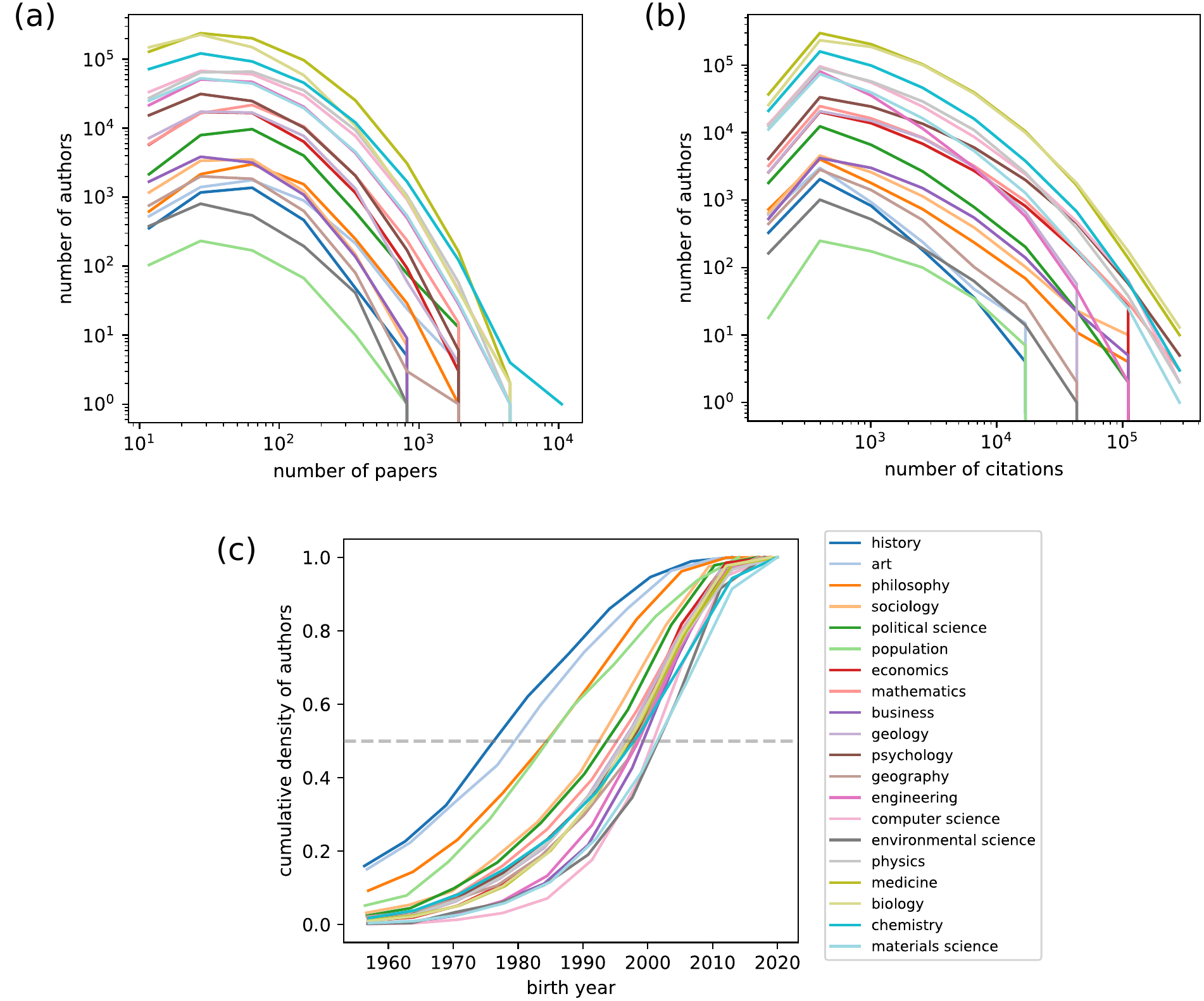}
  \caption{Basic author statistics. Each curve represents authors in a specific field. {Gray dashed horizontal lines indicates the median}. \textbf{(a)} Distribution of number of papers. \textbf{(b)} Distribution of citation counts. \textbf{(c)} Cumulative distribution of authors birth year. 
  }
  \label{fig:r1}
\end{figure}

Figure \ref{fig:r1} shows some basic statistics extracted from our database. The shape of author citation distribution is similar across fields (Figure \ref{fig:r1}(a)), with differences in the number of citations received by different areas.
Typically authors from Biology and Medicine are the ones receiving more citations, while other areas are much less cited ({see e.g. History, Art and Geography}). The number of citations received seems to be related with productivity, as shown in Figure \ref{fig:r1}(b). As expected, different areas have different productivity and visibility patterns. 

Figure \ref{fig:r1}(c) shows the cumulative distribution of authors birth year. The birth year here is defined as the year of the author’s first publication. We show as a dashed horizontal line the median value. Once again the differences across fields are evident, even though shapes are similar. Apart from History, Art, Philosophy and Population, all other subfields have authors with median birth year between 1990 and 2000. Surprisingly, a difference of almost 30 years is observed when comparing the median average birth year of Chemistry and History authors. 

\begin{figure}[h]
    \centering
    \includegraphics[width=\textwidth]{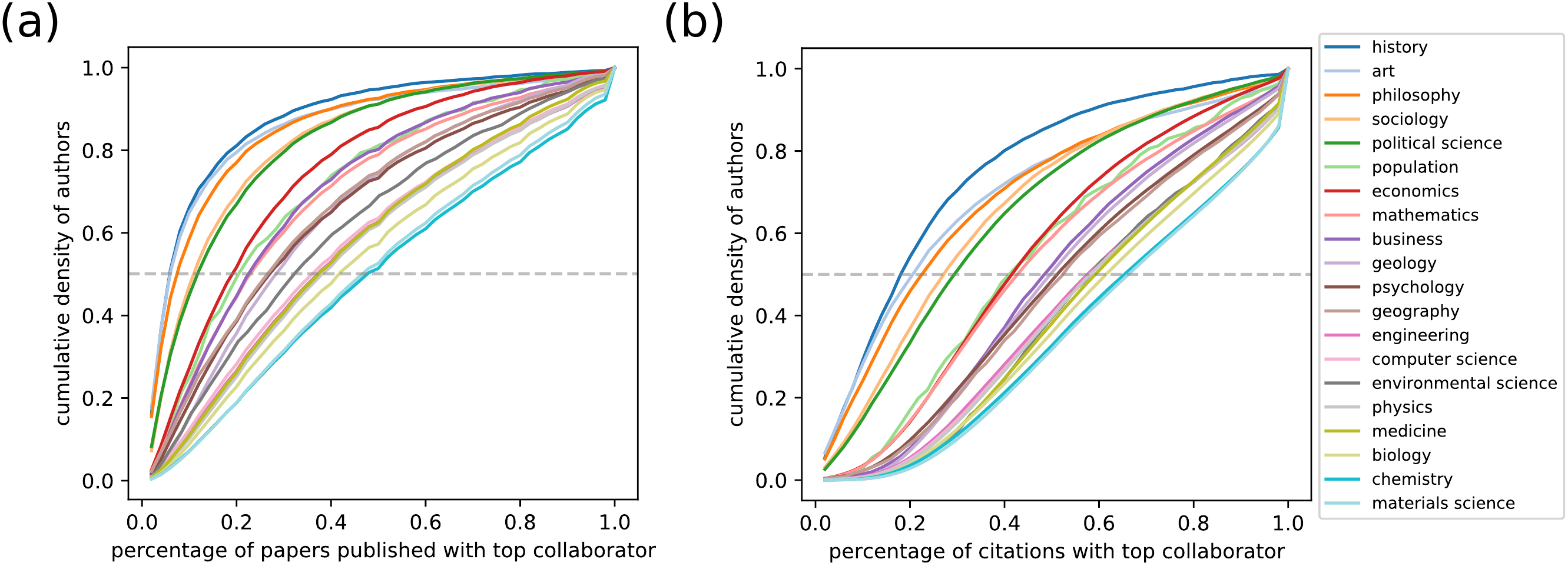}
    \caption{Cumulative distributions of top-collaborators influence considering in different fields of study. The grey line indicates the median. The top-collaborator influence is measured in terms of (a) number of papers; and (b) citation counts. }
    \label{fig:sub4}
\end{figure}

We investigated the influence of the top-collaborator (as defined in equation \ref{eq:topa}) in different subfields. We analyzed the weight associated with the top-collaborator of A (i.e. $\max_X w_{AX}$).
{The median values of the distribution of $\max_X w_{AX}$ when defining weights in terms of the publication citation counts is shown in Table \ref{tab:my_label} of the SI.}  
In Figure \ref{fig:sub4}(a), we show the cumulative distribution of authors for $\max_X w_{AX}$, where weights are defined in terms of authors productivity. Equivalently, the figure illustrates  the cumulative distribution of the fraction of papers published with the top collaborator. Interestingly, there are significant differences across fields. The median top collaborator influence (denoted as dashed horizontal line) ranges between 7\% and can reach roughly 50\%.  History, Art, Philosophy and Sociology are the areas displaying lowest influence of top collaborators in publication counts. Conversely, Material Sciences, Chemistry, Biology and Medicine are the areas with a high influence of top collaborators. For instance, for half of all Chemistry authors, top collaborators accounts for roughly 50\% of publication counts. 

Figure \ref{fig:sub4}(b) shows the cumulative density of top-collaborators influence when considering citation counts. The shapes are very similar to the ones found for publication counts. However, when considering citations, the influence of top-collaborators seems to be stronger. The median influence can increase by a margin of $20\%$ in some disciplines. In Material Sciences, the median influence based on citation reaches $66\%$.

We note that there is no trivial correlation between the features of areas (as shown in Figure \ref{fig:r1}) and influence indexes found in Figure \ref{fig:sub4}. While Biology and Medicine are the disciplines with the largest number of papers and citations, the top-collaborator seems to be more influential in both Chemistry and Material Sciences. Similarly, Art is the area with the lowest number of papers and citations, however, the lowest influence of top-collaborators is observed in History. 
We could also observe that authors' birth year can not totally explain top-collaborators influence. While Computer Science seems to be one of the youngest fields, the median dependence is much higher in fields such as Biology and Chemistry. These results suggest that the influence of top-collaborators might be related to additional factors other than productivity/citation behavior or authors age distribution.  

An alternative way to investigate the influence of top-collaborators concerns the analysis of how top-collaborators can affect authors visibility metrics. In this context, we analyzed the variation of authors visibility indexes when the contribution of the top-collaborator is removed. In this analysis, we should expect that if a single top collaborator accounts for most of the success of an author, if the contribution (i.e. co-authored papers) of that collaborator is disregarded, the visibility of the author is affected by a large margin. Conversely, if collaborations are diverse in quantity and quality, one should expect that disregarding top-collaborators contributions would cause only a minor decrease in the considered visibility indexes. 

\begin{figure}
    \centering
    \includegraphics[width=0.95\textwidth]{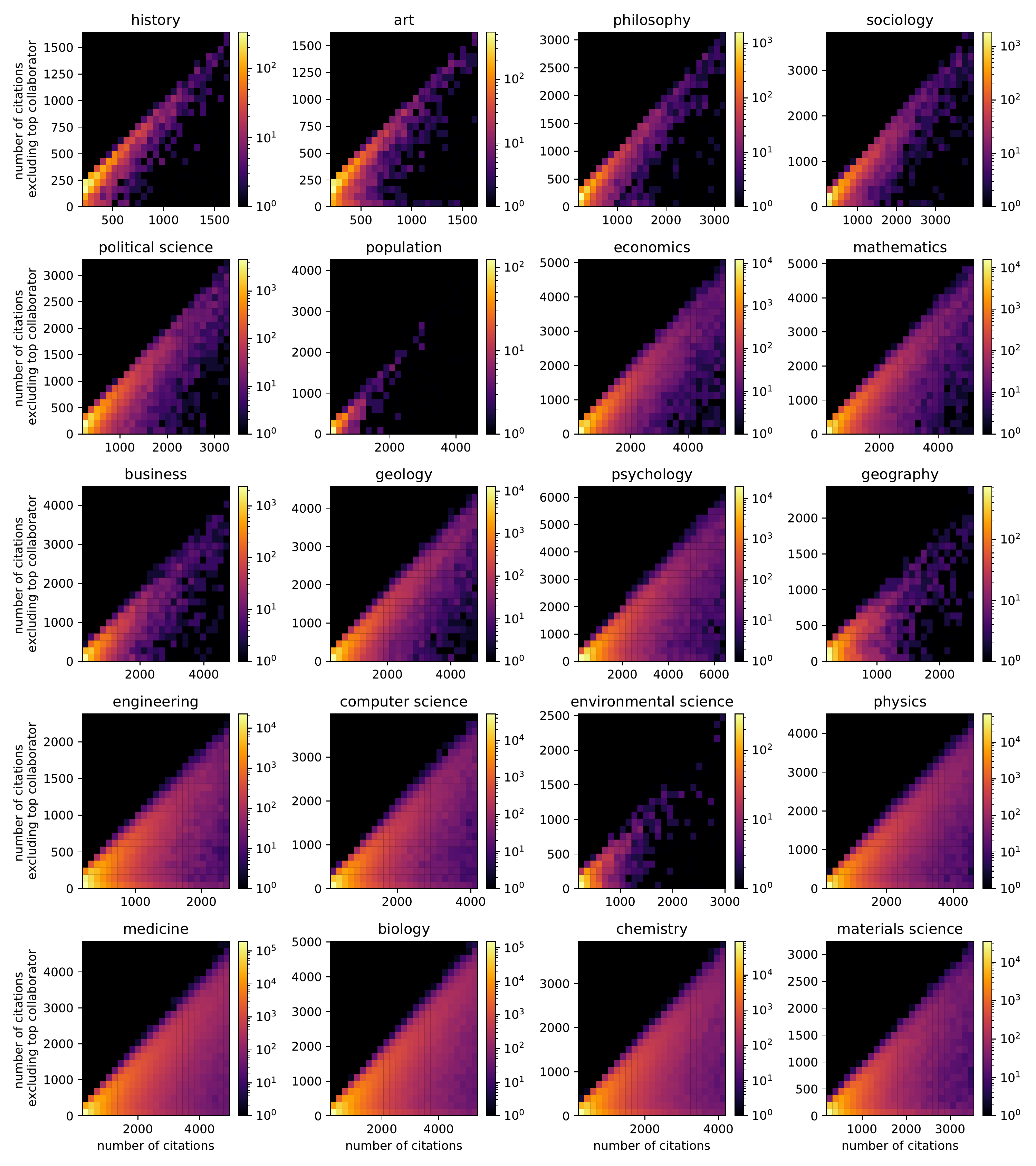}
    \caption{Total number of citations (x-axis) vs. citations obtained disregarding papers co-authored with top-collaborators (y-axis). While in History, Art, Philosophy and Sociology top-collaborators  account for a small percentage of citations, in other disciplines such as Medicine and Chemistry, top-collaborators may play an important individual role to authors citation counts.
    }
    \label{fig:cits}
\end{figure}

In Figure \ref{fig:cits} we analyze the robustness of authors metrics when the contribution of top-collaborators are disregarded. More specifically, we analyze how \emph{citation counts} are affected when papers with the top-collaborators are disregarded from the analysis. {A similar study analyzing the effect on \emph{publication counts} is available in Figure \ref{fig:papers} of the SI.}
For each subpanel, the x-axis corresponds to the original author citation counts, and the y-axis denotes the number of citations accrued by authors in papers \emph{not co-authored with their top-collaborators}. While there are noteworthy differences across fields, it is possible to identify that, for authors with low citations, removing the top-collaborator can strongly affect citation performance. This is not a surprising result, since this might be the case of young researchers who mostly collaborated with their supervisors. This effect might not only be restricted to younger researchers, since only a small percentage of authors have birth date before 2010 (as shown in Figure \ref{fig:r1}(c)). In fact, a correlation analysis  between authors age and top-collaborators influence showed that there is no strong relationship between the two variables (see Figure \ref{fig:agecor} of the SI). 


The influence of top-collaborators on citation counts is field dependent. Particularly, a small dependence can be observed in History, Art, Philosophy, Sociology, Population and Environment Sciences. Even for highly cited authors, when removing the top-collaborator, citation counts are not affected by a large margin. Conversely, it is interesting to note that, for some fields, removing the top-collaborator may cause a large impact on author citations. This pattern emerges in several fields, including Engineering, Computer Science, Physics,  Medicine, Biology, Chemistry and Material Sciences. It is also worthy noting that even highly-cited authors can be strongly affected by removing top-collaborators publications. In Medicine, authors with more than $4,000$ citations can share more than $3,000$ citations with their single top-collaborators. In Economics, Mathematics and Psychology the removal of top-collaborators from the analysis can also impact significantly citation counts of highly-cited authors. However, this effect is not as frequent as observed in Medicine and Chemistry. 

\begin{figure}
    \centering
    \includegraphics[width=0.95\textwidth]{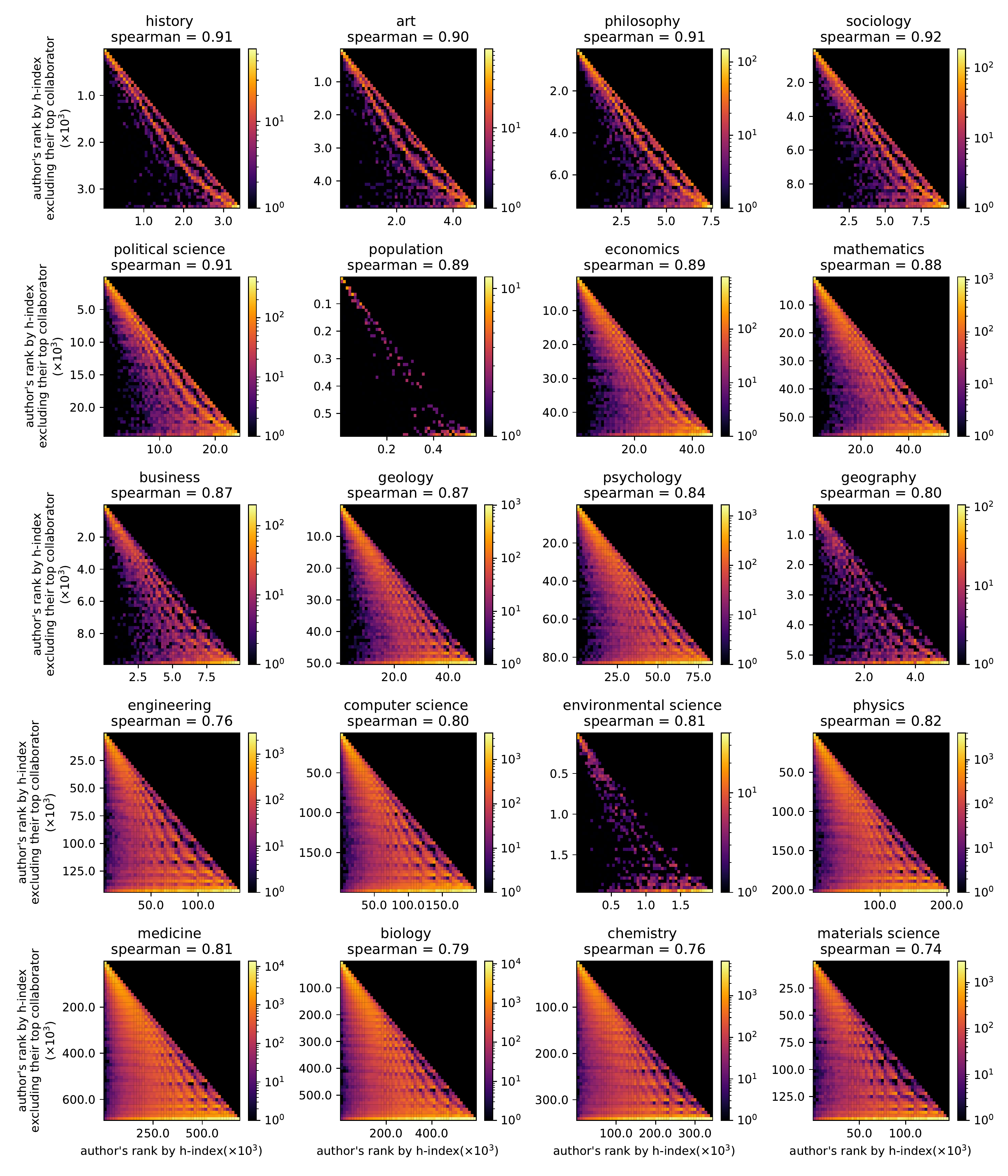}
    \caption{Analysis of how $h$-based authors ranking is affected by top-collaborators contributions. The x-axis represents the ranking obtained with the generalized h-index and the y-axis corresponds to the ranking of the same metrics but disregarding top-collaborator contributions. 
    }
    \label{fig:rankhindex}
\end{figure}

In Figure \ref{fig:rankhindex} we conducted a similar experiment by considering the \emph{generalized h-index} as visibility metric. The x-axis corresponds to authors' rank when sorting author according to their \emph{h-indexes}. The y-axis is the analogous ranking, but disregarding the contributions of top-collaborators. The best ranked authors  are located at the top left region of each subpanel. 
We observe again that the importance in terms of h-indexes is clearly field dependent.
Once again the main differences across areas arises for the most visible (best ranked) researchers. In History, Art, Philosophy, Sociology, Population and Environmental Sciences, the original rank is affected mostly  for the authors with the lowest h-indexes values. The h-indexes of the best ranked authors seems not to be strongly dependent on the contributions of top-collaborators for these disciplines. 

The disciplines mostly dependent on top-collaborators are Engineering, Computer Science, Physics, Medicine, Biology, Chemistry and Materiaks Sciences. For some authors in these areas, their ranking can be strongly impacted if top-collaborators contributions are disregarded. Even top-ranked researchers can be affected by removing top-collaborators papers. This result suggests that a single, recurrent collaboration may play a prominent role in establishing a high visibility level in these disciplines.

\subsection{Relationship between number of co-authors and top-collaborators influence}

The previous section showed that top-collaborators influence based on visibility metrics is field dependent. It becomes thus interesting to investigate whether these patterns can be explained by the differences in the typical number of collaborators in each field. This is an issue worthy studying because, when authors publish several single-authored papers, the lack of top-collaborators can be trivially explained. As we shall show, the predominance of single-authored papers in humanities could be one reason for the low influence of top-collaborators in disciplines such as History, Art and Philosophy~\citep{sabharwal2013comparing}.

\begin{figure}[h]
    \centering
    \includegraphics[width=1\textwidth]{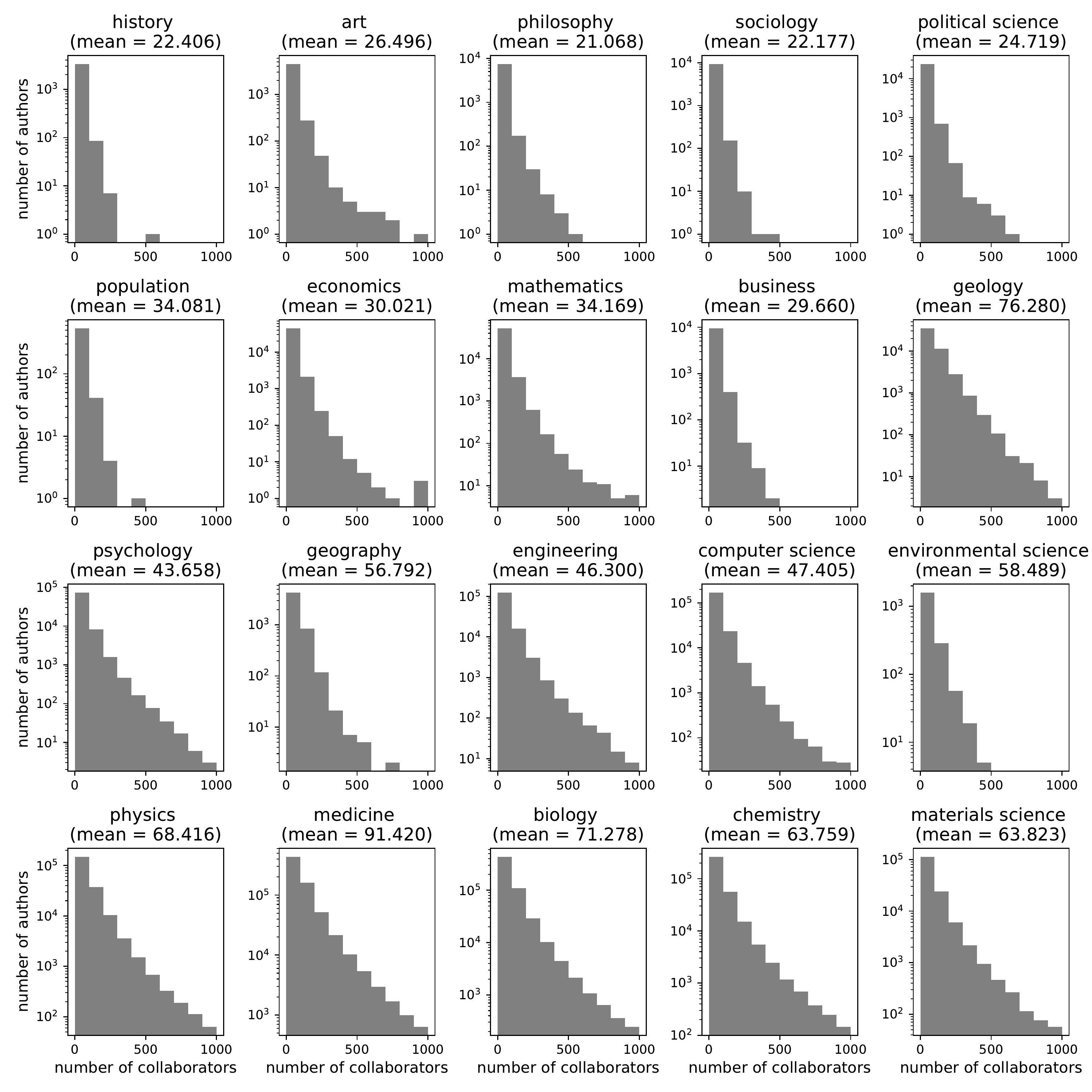}
    \caption{Distribution of number of collaborators per field of study. The average value is calculated ignoring the values greater than the 0.95 percentile. 
    }
    \label{fig:ncolabs}
\end{figure}

Figure \ref{fig:ncolabs} shows the distribution of the number of collaborators for each discipline. We observe that for some disciplines the low influence of top-collaborators may be correlated with the typical low number of collaborators. This is exactly the case of humanities disciplines. A large dependence of top-collaborators also might be correlated with a large number of collaborations, and this is the case of Medicine and Biology. However, not all influence patterns can be explained totally by the number of collaborations. Geology and Psychology have similar top-collaborator influence profiles (see Figure \ref{fig:rankhindex}). However, the typical number of collaborators are very distinct: Geology has roughly 75\% more collaborators than Psychology. In a similar fashion, Geology and Biology share a similar typical number of collaborators, but top-collaborators influence is clearly stronger in the Biology field. 



\begin{figure}[h]
    \centering
    \includegraphics[width=0.7\textwidth]{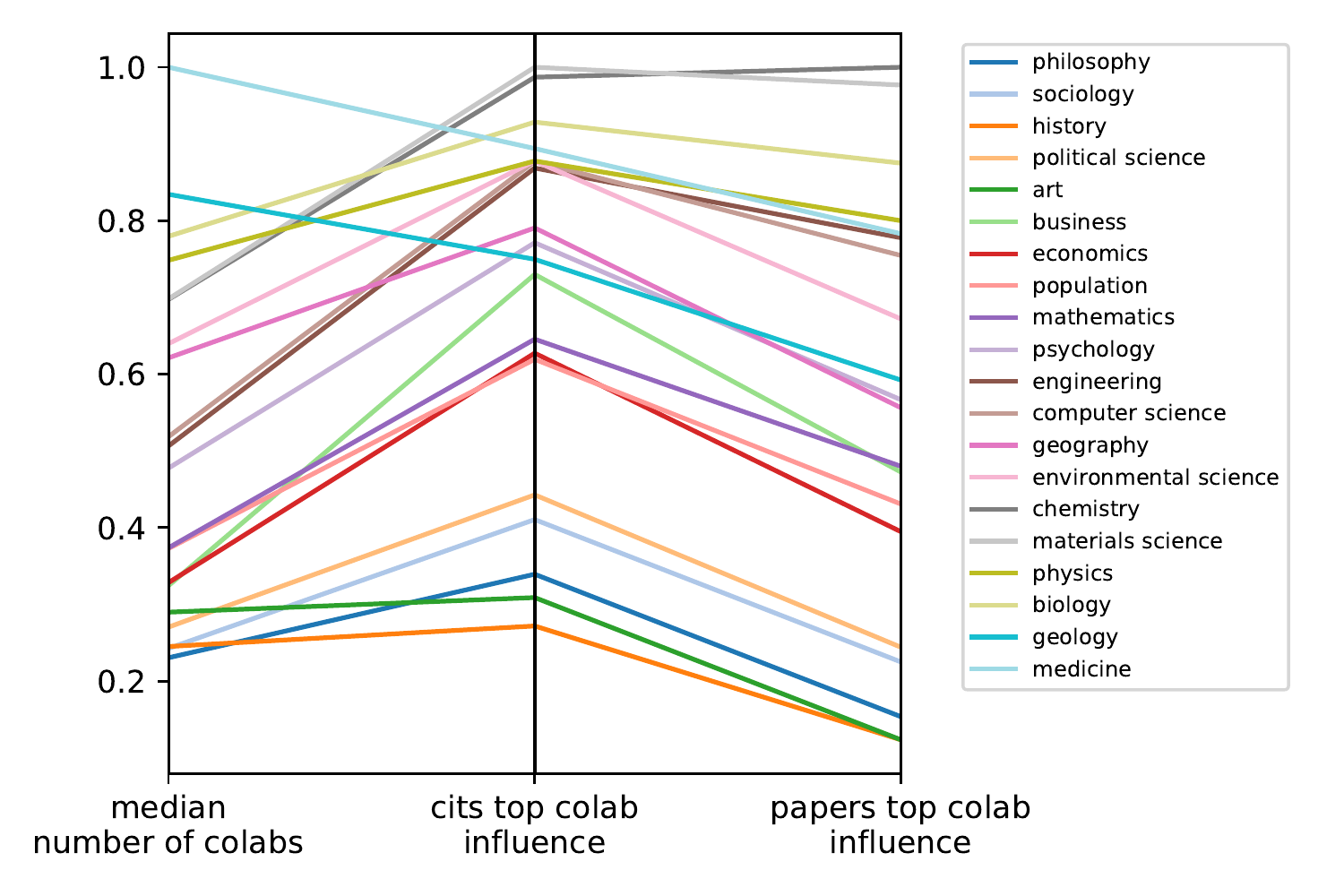}
    \caption{Parallel coordinates illustrating the relationship between the number of collaborators, the influence of top-collaborators on citations and the influence of  top-collaborators on publication counts. Metrics are normalized by the maximum quantity in each axis. }
    \label{fig:comparison}
\end{figure}

Figure \ref{fig:comparison} ranks the disciplines according to different authors' metrics: the average number of collaborators and the median of the influence of the top-collaborator on citations and number of papers. The values in each each axis is normalized by the maximum value observed for the specific variable. The non-normalized median values of top-collaborators influence are displayed in Table \ref{tab:my_label} of the SI. The figure confirms that the relationship between top-collaborators influence and the number of collaborators is not trivial. Medicine is the discipline with the highest number of collaborations, but the influence of top-collaborators is not as high as in other disciplines, such as Material Sciences and Chemistry. The curve observed for Geology is similar to the Medicine curve. An interesting behavior is also observed for Business. Despite being an area with a relative low number of collaborators, the average top-collaborator influence on citations is high. 

All in all our analysis showed that more collaborative disciplines are more likely to have top-collaborators with a higher level of influence on authors productivity and visibility. In some cases, however, top-collaborators influence are relevant even in disciplines characterized by a lower number of collaborators. It remains to be shown, therefore, that other factors other than the number of collaborators may affect top-collaborators influence. 



\section{Conclusions}
\label{sec:conc}

In scientific research, collaborations play a fundamental role in promoting and spreading diverse ideas. Despite the large number of works conducted to study collaboration networks, the influence of a single collaborator on authors productivity and visibility has been limited to a few works. In this paper, we analyzed the influence of the most relevant collaborators (referred to as \emph{top-collaborator}) on author metrics. The influence of top-collaborators on the number of papers, citations and h-index was studied in several disciplines. 

The influence of the top-collaborator was defined as the fraction of papers/citations obtained in studies co-authored with the top-collaborator. Several interesting patterns distinguishing disciplines were found. We identified that disciplines with a typical low number of collaborators (such as Philosophy, History and Art) are more likely to have a high top-collaborator importance. In this sense, the number of papers and citations are not highly affected if top-collaborators contributions are not affected. A second group of discipline are formed by authors with more collaborators and moderate influence of the top collaborators. This group includes e.g. Economics and Mathematics. A third group of disciplines are those displaying an inversely proportional relationship between number of collaborators and influence of the top-collaborator. Medicine and Geology are the two most collaborative disciplines, however, the influence of top-collaborators is surpassed by other disciplines. In an opposite behavior, Business is a typically a discipline with a relative low number of collaborations, but the influence of top-collaborators in citations is roughly the same the influence observed for Geology, which is a much more collaborative discipline. 


Our results showed that the top-collaborator may play an important role in particular disciplines even for highly cited authors. This implies disentangling individual's success from collaborative their teams has become a challenging task and will get more complex as science becomes even more collaborative. In addition to that, we found different patterns of collaboration influence among different disciplines. Thus, any new metric to measure individual research impact should be versatile enough to account for the specificities of each discipline, which also poses as a challenging as science has become more interdisiplinary over the past years.

In future works, we are interested in using top-collaborator influence metrics to investigate individual careers. We are particularly interested in investigating if collaboration patterns with top-collaborators could predict a higher future visibility or other relevant author metrics~\citep{brito2021associations}. 
In this case, we could use the proposed network representation to measure the effective number of collaborators via accessibility or symmetry metrics~\citep{correa2017patterns,tohalino2018extractive}. Another interesting topic worthy studying is the contribution performed by top-collaborators~\citep{correa2017patterns}. In other words, we could investigate, e.g. if top-collaborators are more likely to contribute in a specific task, such as writing or research design. 


\newpage

\section*{Acknowledgments}

A.C.M.B. acknowledges financial support from São Paulo Research Foundation (FAPESP Grant no. 2020/14817-2) and Capes-Brazil for sponsorship. D.R.A. acknowledges financial support from FAPESP (grant no. 20/06271-0) and CNPq-Brazil (grant no. 304026/2018-2). 

\bibliographystyle{abbrvnat}
\bibliography{main}

\begin{thebibliography}{26}
\providecommand{\natexlab}[1]{#1}
\providecommand{\url}[1]{\texttt{#1}}
\expandafter\ifx\csname urlstyle\endcsname\relax
  \providecommand{\doi}[1]{doi: #1}\else
  \providecommand{\doi}{doi: \begingroup \urlstyle{rm}\Url}\fi

\bibitem[Abramo and D’Angelo(2014)]{abramo2014you}
G.~Abramo and C.~A. D’Angelo.
\newblock How do you define and measure research productivity?
\newblock \emph{Scientometrics}, 101\penalty0 (2):\penalty0 1129--1144, 2014.

\bibitem[Abramo and D’Angelo(2015)]{abramo2015relationship}
G.~Abramo and C.~A. D’Angelo.
\newblock The relationship between the number of authors of a publication, its
  citations and the impact factor of the publishing journal: Evidence from
  italy.
\newblock \emph{Journal of Informetrics}, 9\penalty0 (4):\penalty0 746--761,
  2015.

\bibitem[Abramo et~al.(2009)Abramo, D’Angelo, and
  Di~Costa]{abramo2009research}
G.~Abramo, C.~A. D’Angelo, and F.~Di~Costa.
\newblock Research collaboration and productivity: is there correlation?
\newblock \emph{Higher education}, 57\penalty0 (2):\penalty0 155--171, 2009.

\bibitem[Abramo et~al.(2013)Abramo, Cicero, and
  D’Angelo]{abramo2013individual}
G.~Abramo, T.~Cicero, and C.~A. D’Angelo.
\newblock Individual research performance: A proposal for comparing apples to
  oranges.
\newblock \emph{Journal of Informetrics}, 7\penalty0 (2):\penalty0 528--539,
  2013.

\bibitem[Ajiferuke and Wolfram(2010)]{ajiferuke2010citer}
I.~Ajiferuke and D.~Wolfram.
\newblock Citer analysis as a measure of research impact: Library and
  information science as a case study.
\newblock \emph{Scientometrics}, 83\penalty0 (3):\penalty0 623--638, 2010.

\bibitem[Ajiferuke et~al.(2010)Ajiferuke, Lu, and
  Wolfram]{ajiferuke2010comparison}
I.~Ajiferuke, K.~Lu, and D.~Wolfram.
\newblock A comparison of citer and citation-based measure outcomes for
  multiple disciplines.
\newblock \emph{Journal of the American Society for Information Science and
  technology}, 61\penalty0 (10):\penalty0 2086--2096, 2010.

\bibitem[Amancio et~al.(2012)Amancio, Oliveira~Jr, and Costa]{amancio2012use}
D.~R. Amancio, O.~N. Oliveira~Jr, and L.~d.~F. Costa.
\newblock On the use of topological features and hierarchical characterization
  for disambiguating names in collaborative networks.
\newblock \emph{EPL (Europhysics Letters)}, 99\penalty0 (4):\penalty0 48002,
  2012.

\bibitem[Amancio et~al.(2015)Amancio, Oliveira~Jr, and
  Costa]{amancio2015topological}
D.~R. Amancio, O.~N. Oliveira~Jr, and L.~d.~F. Costa.
\newblock Topological-collaborative approach for disambiguating authors’
  names in collaborative networks.
\newblock \emph{Scientometrics}, 102\penalty0 (1):\penalty0 465--485, 2015.

\bibitem[Batista et~al.(2006)Batista, Campiteli, and
  Kinouchi]{batista2006possible}
P.~D. Batista, M.~G. Campiteli, and O.~Kinouchi.
\newblock Is it possible to compare researchers with different scientific
  interests?
\newblock \emph{Scientometrics}, 68\penalty0 (1):\penalty0 179--189, 2006.

\bibitem[Beaver(2001)]{beaver2001reflections}
D.~Beaver.
\newblock Reflections on scientific collaboration (and its study): past,
  present, and future.
\newblock \emph{Scientometrics}, 52\penalty0 (3):\penalty0 365--377, 2001.

\bibitem[Brito et~al.(2021)Brito, Silva, and Amancio]{brito2021associations}
A.~C. Brito, F.~N. Silva, and D.~R. Amancio.
\newblock Associations between author-level metrics in subsequent time periods.
\newblock \emph{Journal of Informetrics}, 15\penalty0 (4):\penalty0 101218,
  2021.

\bibitem[Bukvova(2010)]{bukvova2010studying}
H.~Bukvova.
\newblock Studying research collaboration: A literature review.
\newblock 2010.

\bibitem[Corr{\^e}a~Jr et~al.(2017)Corr{\^e}a~Jr, Silva, Costa, and
  Amancio]{correa2017patterns}
E.~A. Corr{\^e}a~Jr, F.~N. Silva, L.~d.~F. Costa, and D.~R. Amancio.
\newblock Patterns of authors contribution in scientific manuscripts.
\newblock \emph{Journal of Informetrics}, 11\penalty0 (2):\penalty0 498--510,
  2017.

\bibitem[Fenner(2014)]{fenner2014altmetrics}
M.~Fenner.
\newblock Altmetrics and other novel measures for scientific impact.
\newblock In \emph{Opening science}, pages 179--189. Springer, Cham, 2014.

\bibitem[Freeman et~al.(2015)Freeman, Ganguli, and
  Murciano-Goroff]{freeman20151}
R.~B. Freeman, I.~Ganguli, and R.~Murciano-Goroff.
\newblock \emph{Why and Wherefore of Increased Scientific Collaboration}.
\newblock University of Chicago Press, 2015.

\bibitem[Garc{\'\i}a-P{\'e}rez(2012)]{garcia2012extension}
M.~A. Garc{\'\i}a-P{\'e}rez.
\newblock An extension of the h index that covers the tail and the top of the
  citation curve and allows ranking researchers with similar h.
\newblock \emph{Journal of Informetrics}, 6\penalty0 (4):\penalty0 689--699,
  2012.

\bibitem[Ioannidis et~al.(2016)Ioannidis, Klavans, and
  Boyack]{ioannidis2016multiple}
J.~P. Ioannidis, R.~Klavans, and K.~W. Boyack.
\newblock Multiple citation indicators and their composite across scientific
  disciplines.
\newblock \emph{PLoS biology}, 14\penalty0 (7):\penalty0 e1002501, 2016.

\bibitem[Ioannidis et~al.(2020)Ioannidis, Boyack, and
  Baas]{ioannidis2020updated}
J.~P. Ioannidis, K.~W. Boyack, and J.~Baas.
\newblock Updated science-wide author databases of standardized citation
  indicators.
\newblock \emph{Plos Biology}, 18\penalty0 (10):\penalty0 e3000918, 2020.

\bibitem[Larivi{\`e}re et~al.(2015)Larivi{\`e}re, Gingras, Sugimoto, and
  Tsou]{lariviere2015team}
V.~Larivi{\`e}re, Y.~Gingras, C.~R. Sugimoto, and A.~Tsou.
\newblock Team size matters: Collaboration and scientific impact since 1900.
\newblock \emph{Journal of the Association for Information Science and
  Technology}, 66\penalty0 (7):\penalty0 1323--1332, 2015.

\bibitem[Li et~al.(2019)Li, Aste, Caccioli, and Livan]{li2019reciprocity}
W.~Li, T.~Aste, F.~Caccioli, and G.~Livan.
\newblock Reciprocity and impact in academic careers.
\newblock \emph{EPJ Data Science}, 8\penalty0 (1):\penalty0 20, 2019.

\bibitem[Pelacho et~al.(2021)Pelacho, Ruiz, Sanz, Taranc{\'o}n, and
  Clemente-Gallardo]{pelacho2021analysis}
M.~Pelacho, G.~Ruiz, F.~Sanz, A.~Taranc{\'o}n, and J.~Clemente-Gallardo.
\newblock Analysis of the evolution and collaboration networks of citizen
  science scientific publications.
\newblock \emph{Scientometrics}, 126\penalty0 (1):\penalty0 225--257, 2021.

\bibitem[Qin et~al.(1997)Qin, Lancaster, and Allen]{qin1997types}
J.~Qin, F.~W. Lancaster, and B.~Allen.
\newblock Types and levels of collaboration in interdisciplinary research in
  the sciences.
\newblock \emph{Journal of the American Society for information Science},
  48\penalty0 (10):\penalty0 893--916, 1997.

\bibitem[Sabharwal(2013)]{sabharwal2013comparing}
M.~Sabharwal.
\newblock Comparing research productivity across disciplines and career stages.
\newblock \emph{Journal of Comparative Policy Analysis: Research and Practice},
  15\penalty0 (2):\penalty0 141--163, 2013.

\bibitem[Tohalino and Amancio(2018)]{tohalino2018extractive}
J.~V. Tohalino and D.~R. Amancio.
\newblock Extractive multi-document summarization using multilayer networks.
\newblock \emph{Physica A: Statistical Mechanics and its Applications},
  503:\penalty0 526--539, 2018.

\bibitem[Viana et~al.(2013)Viana, Amancio, and Costa]{viana2013time}
M.~P. Viana, D.~R. Amancio, and L.~d.~F. Costa.
\newblock On time-varying collaboration networks.
\newblock \emph{Journal of Informetrics}, 7\penalty0 (2):\penalty0 371--378,
  2013.

\bibitem[Wang et~al.(2020)Wang, Shen, Huang, Wu, Dong, and
  Kanakia]{wang2020microsoft}
K.~Wang, Z.~Shen, C.~Huang, C.-H. Wu, Y.~Dong, and A.~Kanakia.
\newblock Microsoft academic graph: When experts are not enough.
\newblock \emph{Quantitative Science Studies}, 1\penalty0 (1):\penalty0
  396--413, 2020.

\end{thebibliography}

\newpage

\section*{Supplementary Information}

\setcounter{figure}{0}
\renewcommand{\figurename}{FIG.}
\renewcommand{\thefigure}{S\arabic{figure}}

\setcounter{table}{0}
\renewcommand{\tablename}{TABLE}
\renewcommand{\thetable}{S\arabic{table}}

\begin{figure}[ht!]
    \centering
    \includegraphics[width=0.60\textwidth]{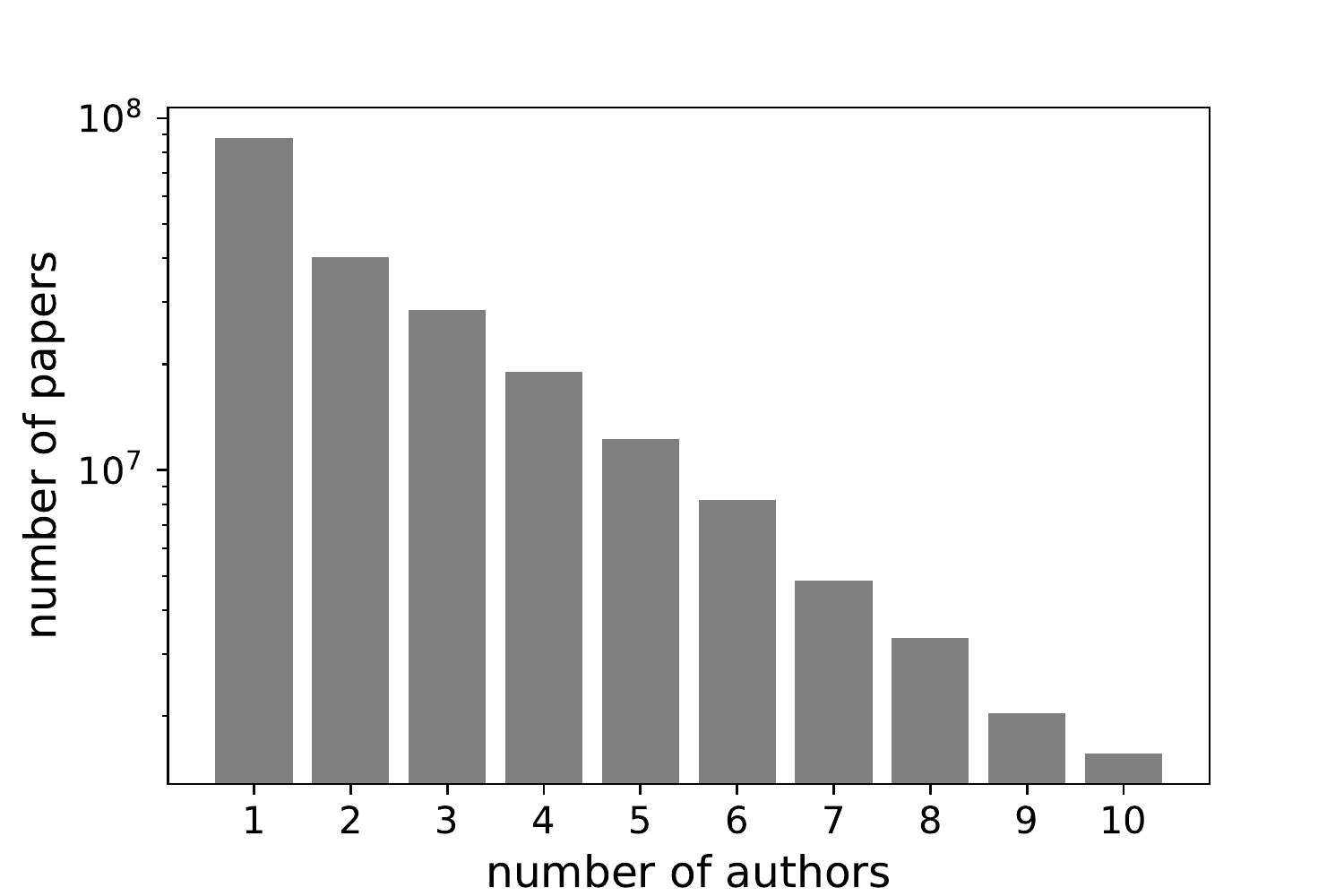}
    \caption{Distribution of the number of authors per publications (1950 -- 2020). 
    }
    \label{fig:num-authors}
\end{figure}

\begin{table}[h]
\centering
    \caption{Median top-collaborator influence taking into account the number of papers and the number of citations. The values for each column are sorted in increasing order. }
\begin{tabular}{|l|c|l|c|}
\hline
                {\bf Discipline} &  {\bf Papers} & {\bf Discipline} & {\bf Citations} \\
\hline
              History &                            0.06 &               History &                               0.18 \\
                  Art &                            0.06 &                   Art &                               0.20 \\
           Philosophy &                            0.07 &            Philosophy &                               0.22 \\
            Sociology &                            0.11 &             Sociology &                               0.27 \\
    Pol. Science &                            0.12 &     Pol. Science &                               0.29 \\
            Economics &                            0.19 &            Population &                               0.41 \\
           Population &                            0.21 &             Economics &                               0.42 \\
             Business &                            0.23 &           Mathematics &                               0.43 \\
          Mathematics &                            0.23 &              Business &                               0.48 \\
            Geography &                            0.26 &               Geology &                               0.50 \\
           Psychology &                            0.27 &            Psychology &                               0.51 \\
              Geology &                            0.28 &             Geography &                               0.52 \\
Env. Science &                            0.32 &           Engineering &                               0.58 \\
     Comp. Science &                            0.36 &      Comp. Science &                               0.58 \\
          Engineering &                            0.37 & Env. science &                               0.58 \\
             Medicine &                            0.37 &               Physics &                               0.58 \\
              Physics &                            0.38 &              Medicine &                               0.59 \\
              Biology &                            0.42 &               Biology &                               0.61 \\
    Mat. Science &                            0.47 &             Chemistry &                               0.65 \\
            Chemistry &                            0.48 &     Mat. Science &                               0.66 \\
    \hline
    \end{tabular}
    \label{tab:my_label}
\end{table}

\begin{figure}[p]
    \centering
    \includegraphics[width=0.95\textwidth]{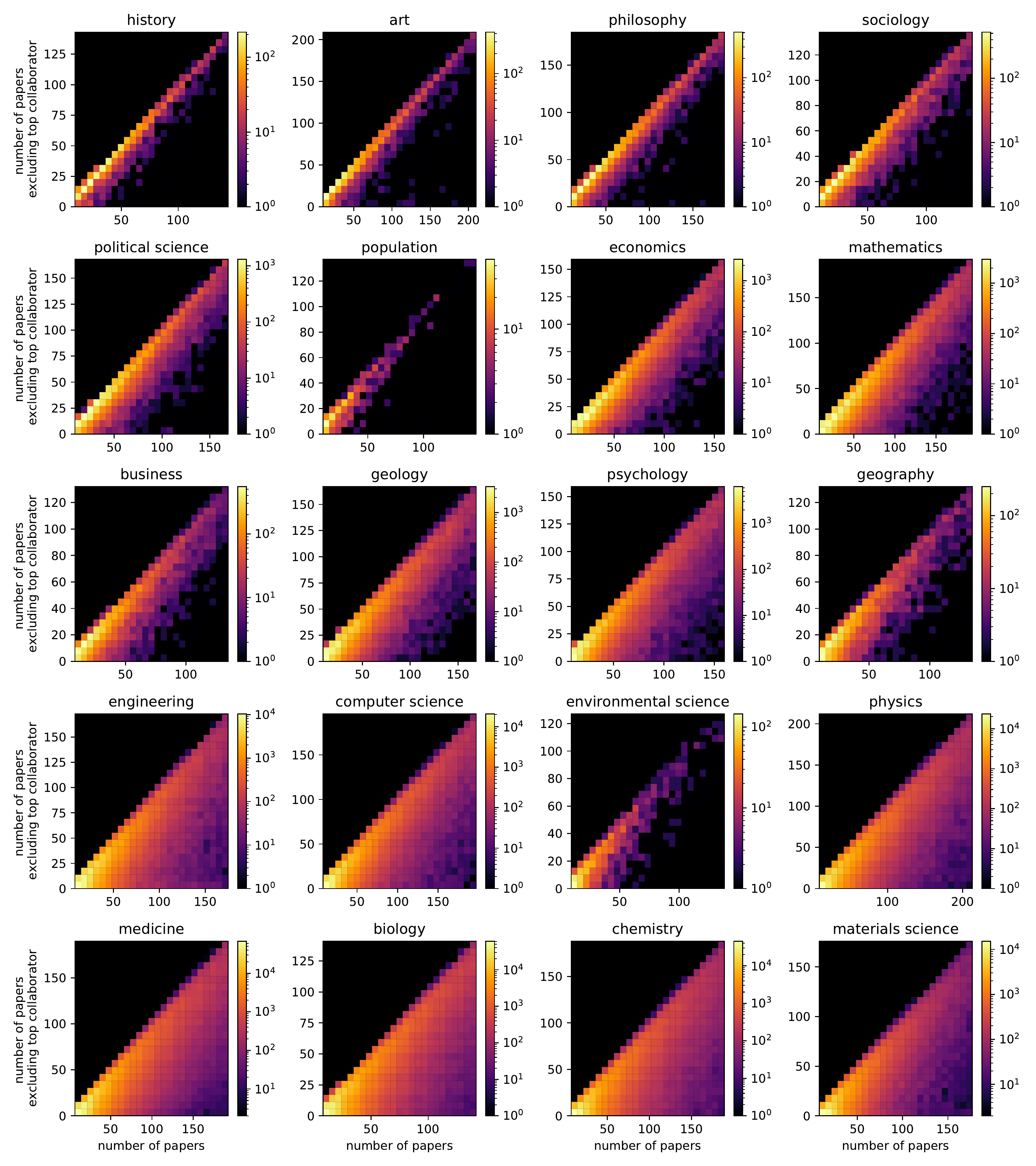}
    \caption{Number of papers considering the top collaborator (x-axis) and ignoring the top collaborator (y-axis). The top-collaborator influence based on paper counts is similar to the one based on citation counts. 
    }
    \label{fig:papers}
\end{figure}

\begin{figure}
    \centering
    \includegraphics[width=0.95\textwidth]{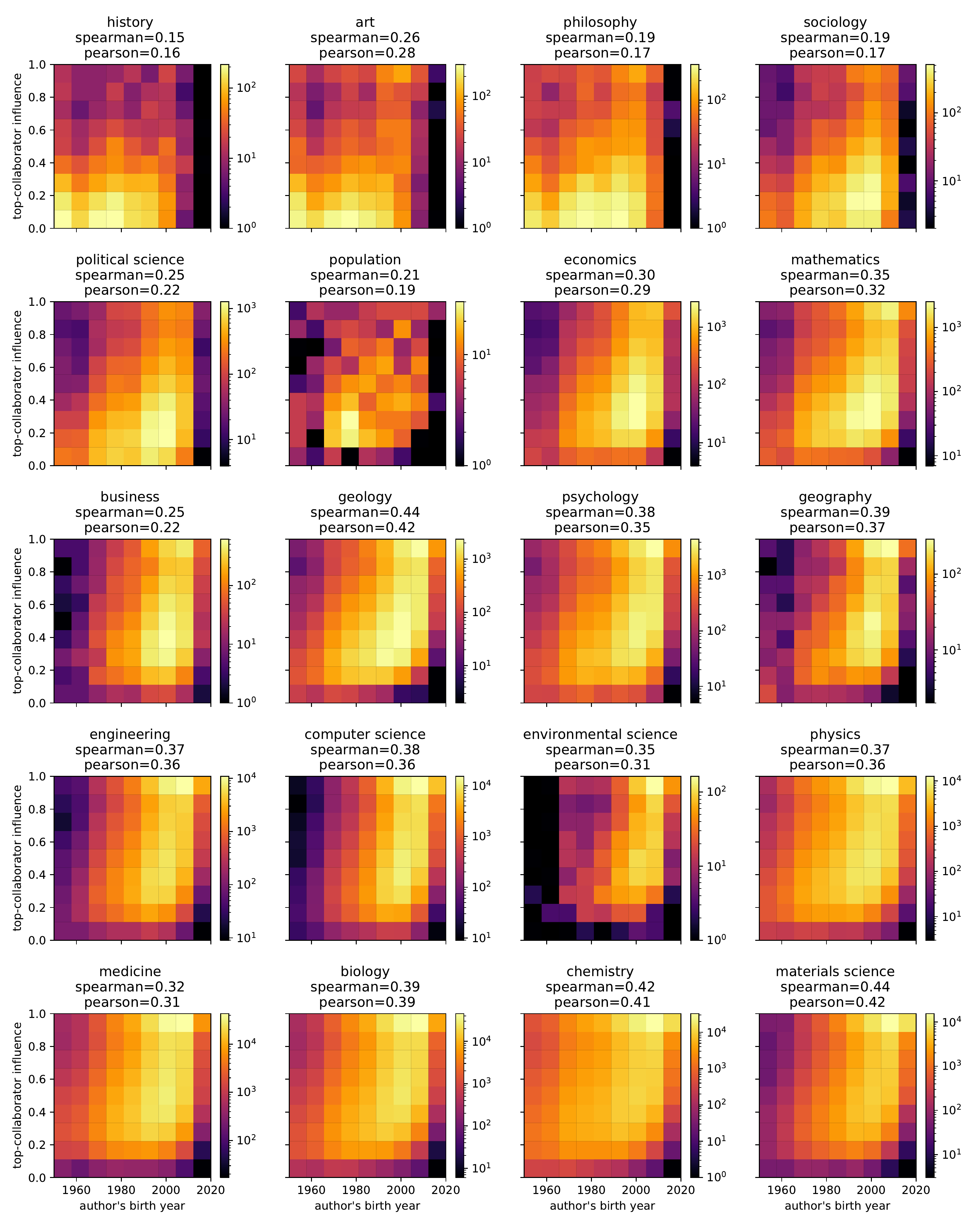}
    \caption{Analysis of the relationship between authors birth year and top-collaborators influence based on citation counts. In all subfields the correlations are either weak or moderate. }
    \label{fig:agecor}
\end{figure}


\end{document}